\documentclass[10pt,conference]{IEEEtran}
\IEEEoverridecommandlockouts
\usepackage{graphicx}
\usepackage{booktabs}
\usepackage{longtable}
\usepackage{hyperref}
\usepackage{pifont}
\usepackage[numbers,compress]{natbib}
\usepackage{multirow}
\usepackage{xspace}
\usepackage{url}
\usepackage{listings}
\usepackage{verbatim}
\usepackage{framed}
\usepackage{graphicx}  
\usepackage{float}  
\usepackage{subfig}    
\usepackage{fancyhdr}
\usepackage[dvipsnames]{xcolor}
\usepackage{algorithm}
\usepackage{algorithmic}
\usepackage{wrapfig}
\usepackage{tcolorbox}
\usepackage{amsmath}
\usepackage{amsthm}
\usepackage{enumitem}

\theoremstyle{definition}
\newtheorem{definition}{Definition}

\lstdefinestyle{cppstyle2}{
  language=C++,                   
  basicstyle=\fontsize{7pt}{8pt}\selectfont\ttfamily,           
  keywordstyle=\color{blue},      
  commentstyle=\color{red!50!green!50!blue!50},     
  stringstyle=\color{red},        
  numbers=none,
  breaklines=true,                
  frame=none,                   
  tabsize=2,                      
  framexleftmargin=1pt,              
  framexrightmargin=1pt,             
  xleftmargin=1pt,                   
  xrightmargin=1pt                   
}

\lstdefinestyle{cppstyle}{
  language=C++,                   
  basicstyle=\tiny\ttfamily,           
  keywordstyle=\color{blue},      
  commentstyle=\color{red!50!green!50!blue!50},     
  stringstyle=\color{red},        
  numbers=none,
  breaklines=true,                
  frame=single,                   
  tabsize=4,                      
  framexleftmargin=1pt,              
  framexrightmargin=1pt,             
  xleftmargin=1pt,                   
  xrightmargin=1pt                   
}

\lstdefinelanguage{json}{
    basicstyle=\tiny\ttfamily, 
    numbers=none,
    stepnumber=1,                      
    numbersep=5pt,                     
    showstringspaces=false,            
    breaklines=true,                   
    frame=single,                      
    literate=
     *{0}{{{\color{blue}0}}}{1}        
      {1}{{{\color{blue}1}}}{1}        
      {2}{{{\color{blue}2}}}{1}        
      {3}{{{\color{blue}3}}}{1}        
      {4}{{{\color{blue}4}}}{1}        
      {5}{{{\color{blue}5}}}{1}        
      {6}{{{\color{blue}6}}}{1}        
      {7}{{{\color{blue}7}}}{1}        
      {8}{{{\color{blue}8}}}{1}        
      {9}{{{\color{blue}9}}}{1}        
      {:}{{{\color{red}:}}}{1}         
      {,}{{{\color{red},}}}{1},        
  framexleftmargin=1pt,              
  framexrightmargin=1pt,             
  xleftmargin=1pt,                   
  xrightmargin=1pt                   
}

\usepackage{soul}	
\setstcolor{red}

\long\def\com#1{}

\newcommand{\kw}[1]{{\it #1}}	
\newcommand{\kn}[1]{\texttt{\small #1}}	
\newcommand{\codet}[1]{{\small \texttt{#1}}}	

\newcommand{\eat}[1]{\if 0 #1 \fi}

\newcommand{\slicer}{\textit{eCPG-Slicer}\xspace}
\newcommand{\mysys}{{LLM4FPM}\xspace}
\newcommand{\myalg}{\textit{FARF}\xspace}
\PassOptionsToPackage{numbers,sort&compress}{natbib}

\def\BibTeX{{\rm B\kern-.05em{\sc i\kern-.025em b}\kern-.08em
    T\kern-.1667em\lower.7ex\hbox{E}\kern-.125emX}}
\begin{document}

\title{Utilizing Precise and Complete Code Context to Guide LLM in Automatic False Positive Mitigation}

\author{
    \IEEEauthorblockN{
        Jinbao Chen,
        Hongjing Xiang,
        Zuohong Zhao,
        Luhao Li,
        Yu Zhang\IEEEauthorrefmark{1}, 
        Boyao Ding,
        Qingwei Li,
        Songyuan Xiong
    }
    \IEEEauthorblockA{
        University of Science and Technology of China, Hefei, Anhui, China \\ 
        Emails: \{zkd18cjb, molepi40, zhaozuohong, llh21\}@mail.ustc.edu.cn, \\ 
        yuzhang@ustc.edu.cn, \{via, lqw332664203, xiongsongyuan\}@mail.ustc.edu.cn
    }
    \thanks{\IEEEauthorrefmark{1}Corresponding author: Yu Zhang (yuzhang@ustc.edu.cn).}
}



\maketitle

\begin{abstract}
Static Application Security Testing (SAST) tools are critical to software quality, identifying potential code issues early in development. 
However, they often produce false positive warnings that require manual review, slowing down development. Thus, automating false positive mitigation (FPM) is essential.
The advent of Large Language Models (LLMs), with their strong abilities in natural language and code understanding, offers promising avenues for FPM. Yet current LLM-based FPM method faces two major limitations: 1. The warning-related code snippets extracted are overly broad and cluttered with irrelevant control/data flows, reducing precision; 2. Critical code contexts are missing, leading to incomplete representations that can mislead LLMs and cause inaccurate assessments. 
To overcome these limitations, we propose \mysys{} , a lightweight and efficient false positive mitigation framework. It features \slicer{}, which builds an extended code property graph (eCPG) to extract precise line-level code contexts for warnings. Furthermore, the integrated \myalg{} algorithm builds a file reference graph to identify all files that are relevant to warnings in linear time. This enables \slicer{} to obtain rich contextual information without resorting to expensive whole-program analysis. 
\mysys{} outperforms the existing method on the Juliet dataset (F1 $>$ 99\% across various Common Weakness Enumerations) and improves label accuracy on the D2A dataset to 86\%. By leveraging a lightweight open-source LLM, \mysys{} 
can significantly save inspection costs up to \$2758 per run (\$0.384 per warning) on Juliet with an average inspection time of 4.7s per warning. Moreover, real-world tests on popular C/C++ projects demonstrate its practicality. 
\end{abstract}




% ...

\begin{figure*}[!t]
  \centering
  \begin{minipage}[b]{.48\textwidth}
  \centering
    \scriptsize
    \begin{minipage}[b]{.48\textwidth}
      \begin{lstlisting}[language=c++,style=cppstyle2]
// from acl_read_cb
static void acl_read_cb(int size, void *priv)
{
 struct data *buf = priv;
 if (size > 0) {
  buf->flag += size;
  buf = NULL;
 }
}
/* Caller of acl_read_cb */
29 static void bluetooth_status_cb(int status)
30 {
      \end{lstlisting}
    \end{minipage}\hfill
    \begin{minipage}[b]{.48\textwidth}
      \begin{lstlisting}[language=c++,style=cppstyle2]
33 switch (status) {
34 case 1:
...
43 case 4:
44  printf("USB device configured");
45  /* Start reading */
46  acl_read_cb(0,NULL);
47  break;
...
60 case 9:
...
64 }
65 }
      \end{lstlisting}
    \end{minipage}
    \vspace{-0.2cm}
    \caption{Positive example of CWE476}
    \label{code:lengthy}
  \end{minipage}
  \hfill
  \begin{minipage}[b]{.25\textwidth}
    \scriptsize
    \centering
    \begin{lstlisting}[language=c++,style=cppstyle2]
// a.c
void CWE369_bad()
{
 float data = 0.0F;
 ...//some flow
 CWE369_badSink(&data);
}

// b.c
void CWE369_badSink(float *dataPtr)
{
 float data = *dataPtr;
 int result = (int)(100.0/data);
}
    \end{lstlisting}
    \vspace{-0.2cm}
    \caption{CWE369 buggy code}
    \label{code:eg-foergincall}
  \end{minipage}
  \hfill
  \begin{minipage}[b]{.23\textwidth}
    \scriptsize
    \centering
    \begin{lstlisting}[language=c++,style=cppstyle2]
// io.c
const int GLOBAL_CONST_TRUE = 1;
// good.c
void CWE121_good() {
 char BadBuffer[50];
 char GoodBuffer[100];
 char *data = BadBuffer;
 if(GLOBAL_CONST_TRUE) {
  data = GoodBuffer;
 }
 char source[100];
 for (size_t i = 0; i < 100; i++) {
  data[i] = source[i];
 }
}
    \end{lstlisting}
    \vspace{-0.2cm}
    \caption{CWE121 good code}
    \label{code:eg-externvar}
  \end{minipage}

  \vspace{-0.3cm}
\end{figure*}

\section{Introduction}








Static Application Security Testing (SAST) tools are indispensable in modern software development, providing automated analysis to detect potential bugs, security vulnerabilities, and other safety or security issues. Applications of these tools include bug detection~\cite{vassallo2020developers, habib2018many}, security code review~\cite{charoenwet2024empirical}, and other activities promoting code safety and security~\cite{chess2004static, livshits2005finding, wei2018amandroid}. 
Companies like Facebook~\cite{distefano2019scaling} and Google~\cite{sadowski2018lessons} have scaled the use of SAST tools across their massive code bases, demonstrating their practicality and value in real-world applications.

However, SAST tools often produce numerous bug reports, with many being false positives~\cite{aloraini2019empirical, yang2019towards}
, leading to developer frustration, wasted time, and resources. 
While SAST warnings are crucial for code safety and security, their high false positive rate undermines their utility and erodes developer trust. As a result, developers may abandon these tools~\cite{johnson2013don}, increasing the risk of overlooked vulnerabilities. \citet{johnson2013don, christakis2016developers} highlight false positives as a major issue for developers and recommend keeping the false positive rate below 20\%.


Manual review of false positives is time-consuming and resource-intensive, especially in large projects. Automatic false positives mitigation (FPM) is thus key to improving SAST tools. Researchers have explored traditional dynamic and static analysis techniques to mitigate false positives, such as combining dependency flow~\cite{muske2019reducing, livshits2009merlin}, program slicing~\cite{rival2005understanding, rival2005abstract} and code tainting~\cite{baca2010identifying}. 
Some studies used machine learning to train classifiers for FPM~\cite{koc2017learning, ranking2014finding}. The advent of LLM, with strong capabilities in natural language processing (NLP) and code understanding, brings new potential for code safety/security~\cite{li2023poster, wu2023large, yang2024large,  pearce2023examining, li2024enhancing}, 
and enabling more effective FPM via integrated program analysis. \citet{wen2024automatically} proposed the first LLM-based FPM system by feeding entire function bodies into the model for analysis setting the current state-of-the-art (SOTA). 

However, there are two main limitations to curent LLM-based FPM approach: 
\ding{172} \textit{Imprecision:} Feeding whole function bodies introduces coarse-grained code context. These large, often noisy snippets can bury the few critical elements relevant to a warning in extensive, irrelevant information, overwhelming LLM and leading to inaccurate assessments, longer processing times, and higher token costs.
Existing studies also indicate that LLMs struggle with long inputs and complex control flow code, limiting their effectiveness~\cite{kwan2023m4le, wang2024beyond,anand2024critical, mordechai2024novicode}. \ding{173} \textit{Incompleteness:} The extracted function bodies derived from bug reports may lack essential code contexts, creating gaps in contextual understanding. Existing SAST tools often fail to include auxiliary files containing crucial elements, such as definitions of global variables or external functions. These elements are vital for accurately understanding the control and data flow relevant to a warning,  and their absence can lead to incorrect conclusions. 

Addressing the above limitations requires extracting precise and complete code context for each warning, a task that requires a dedicated code slicer. However, current code slicers face several challenges: \ding{172} 
They often miss critical dependencies, resulting in imprecise code context. \ding{173} Precise mapping from the slices back to the original source code is difficult, yet crucial for LLM-based analysis to align the code slices with bug reports and the original source code. \ding{174} Whole-program slicing is costly, making it impractical for CI/CD pipelines in real-world projects.

To address the limitations and challenges, we propose \mysys{}, a lightweight and efficient FPM framework. It features \slicer{} for line-level precise slicing and \myalg{}, a linear-time algorithm for identifying warning-relevant files. 

\slicer{} first extends \kn{Joern}~\cite{Joern}'s Code Property Graph (CPG)~\cite{yamaguchi2014modeling} into an enhanced version (eCPG) that captures additional dependencies -- such as call, structural, and variable relationships-- missing from standard CPGs. Guided by bug reports as slicing criteria, it extracts relevant nodes from eCPG and leverages rich source-level metadata in CPG nodes to enable precise mapping back to the source code. 

\mysys{}'s dependent file extractor uses \myalg{} to identify all warning-related code files, ensuring \slicer{} receives comprehensive input without relying on costly whole-program analysis. Finally, \mysys{} composes the sliced context and the bug report into a structured input and applies tailored prompting techniques, enabling the LLM to determine whether the report reflects a true positive.

We evaluated \mysys{} on the Juliet dataset~\cite{juliet}, a widely used open-source dataset for assessing SAST tools, containing a large set of labeled buggy and bug-free C/C++ programs. Results indicate that \mysys{} outperforms the existing LLM-Based method, achieving F1 scores above 99\% across various Common Weakness Enumerations (CWEs). We further applied \mysys{} to the D2A dataset~\cite{zheng2021d2a} to improve its noisy label quality, boosting label accuracy from the original 53\% to 86\%. 
On 8 real-world open-source projects, \mysys{} eliminated over 85\% false positives. In addition, ablation studies confirmed the essential roles of both \slicer{} and \myalg{}. 
Moreover, \mysys{} operates at a low cost, with an average inspection time of 4.7 seconds per warning and no reliance on commercial models--saving up to \$2758 per run on the Juliet dataset by using only lightweight open-source LLMs. 


Overall, our contributions can be summarized as follows:

\begin{itemize}[leftmargin=*]
    \item [\ding{172}] We identify key limitations of current LLM-based FPM and challenges of code context extraction (\autoref{sec:bg}).
    \item [\ding{173}] We introduce \mysys{}, a novel system for LLM-based FPM, which enhances context precision and completeness through \slicer{} and \myalg{} algorithm (\autoref{sec:design}).
    \item [\ding{174}] We demonstrate through extensive evaluation that \mysys{} significantly outperforms SOTA on Juliet, substantially enhances D2A's label accuracy, and reduces false positives in real-world projects (\autoref{sec:eval}).
    \item [\ding{175}] We conclude with a discussion of our work's limitations, related work, and a summary of our key conclusions (\autoref{sec:limits}, \autoref{sec:related}, \autoref{sec:conclusion}).
\end{itemize}

\section{Challengs and Motivation}
\label{sec:bg}
In this section, we first outline several key limitations of current LLM-based FPM and then discuss the challenges of extracting code context. These limitations and challenges motivate our work, setting the stage for our proposed approach.



%
\subsection{Limitations of LLM-based FPM}
\label{sec:bg:moti-example}

Current LLM-based FPM methods struggle to extract accurate code context for LLMs, motivating our work on precise and complete warning-specific extraction.

\subsubsection{Imprecise warning-relevant context}
\label{sec:lengthy}

\citet{wen2024automatically}'s LLM4SA was the first complete system to use LLM for FPM. LLM4SA extracts entire function bodies as code snippets, which are too coarse-grained and lack the precise line-level code context needed by LLMs. 
For example, \autoref{code:lengthy} shows a positive example of CWE476 (NULL Pointer Dereference), where the caller contains a lengthy switch-case statement with 10 cases. However, only case 4 is relevant to the actual calling context, while the rest introduce irrelevant noise. 

Extracting entire function bodies, as show in \autoref{code:lengthy}, yields coarse-grained context that leads to longer LLM inputs, increasing both latency and cost. More critically, studies show that LLMs struggle with long input~\cite{kwan2023m4le, wang2024beyond} and code with complex control/data flow~\cite{anand2024critical, mordechai2024novicode}. 
\autoref{code:lengthy} is just a simplified example. 
Real-world programs often exacerbate this
% ...
with additional API calls, loops, and conditional branches, which can overwhelm LLMs and lead to hallucinations or errors.

\subsubsection{Incomplete warning-relevant context}
\label{sec:bug_report}

Real-world code often references external variables and functions defined in other files, which can critically impact control and data flows, thus affecting bug analysis. However, SAST trace information is often incomplete and omits these external definitions, leading to imprecise bug contexts that may mislead LLMs and result in incorrect conclusions. 


For example, \autoref{code:eg-foergincall} shows a division-by-zero bug from the Juliet dataset, where the error occurs in a callee function (\codet{CWE369\_badSink} in \codet{b.c}) invoked by a caller from another file (\codet{CWE369\_bad} in \codet{a.c}). Some SAST tools report only the caller, omitting the callee's definition and resulting in an incomplete context. Without this definition, it becomes unclear how the division-by-zero is triggered, introducing ambiguity. 


A similar issue arises when external variables influence critical control/data flows. As shown in \autoref{code:eg-externvar}, a benign example related to CWE121 (Buffer Overflow), function \codet{CWE121\_good} references an external variable \codet{GLOBAL\_CONST\_TRUE} (in \codet{io.c}). 
This variable determines whether \codet{data} points to a large enough buffer. 
However, the Cppcheck report only includes \codet{CWE121\_good} and its file, omitting the external definition. 
Without this context, it becomes difficult for the LLM to recognize the bug as a false alarm.

\subsection{Challenges of extracting code context}
\label{sec:challenges}
The key to advancing LLM-based FPM lies in extracting precise and complete code context. Code slicing, a program analysis technique that identifies all statements affecting a specific point (slicing criterion)~\cite{tip1994survey}, is well-suited for this purpose. However, the context extraction process introduces several challenges, which directly motivate our design.

\subsubsection*{Challenge-1:Lack of fidelity or accuracy of current slicer}
\label{sec:challenges:slicer}
To support effective LLM-based analysis of SAST warnings, the slicer must meet two key requirements: \ding{172} Given the warning location as the slicing criterion, it should extract only relevant code context while discarding unrelated parts. \ding{173} The generated slice must retain source-level fidelity, including original code layout and essential metadata (e.g., control flow, file paths, line and column numbers). This precise mapping is vital, as SAST reports refer to the original structure and positions. Maintaining such consistency among the slice, the source code, and the SAST report is therefore essential for accurate LLM-based reasoning.

While some static slicers exist, none fully meet our requirements. For instance, Frama-C Slicing\cite{frama-c_slicing} for C programs outputs the transformed code that lacks original layout, code context, line/column numbers and file structure metadata. Similarly, JavaSlicer~\cite{JavaSlicer/978-3-031-17108-6_9} fails to preserve source fidelity and requires all third-party libraries, limiting its practicality. Recent LLM-based slicing methods~\cite{shahandashti2024program} also fall short, with low accuracy on complex control flow, highlighting the need for significant improvements.


Existing code slicers, as previously discussed, do not fully meet our two requirements, thus necessitating a new approach. Inspired by \citet{zheng2021vu1spg}'s use of \kn{Joern}~\cite{Joern} for code representation in vulnerability detection, we considered it as a promising foundation. \kn{Joern}'s Code Property Graph (CPG) is well-suited for our needs as it effectively combines crucial program dependencies with detailed source-level information. Specifically, it incorporates rich program dependencies derived from Abstract Syntax Tree (AST), Control Flow Graph (CFG), and Program Dependence Graph (PDG), which are essential for effective slicing. Additionally, CPG embeds precise source metadata (e.g., line numbers and filenames) as graph attributes.
Therefore, we selected \kn{Joern}'s CPG as the basis of our slicer to meet the two requirements above. Its other inherent capabilities, including multi-language support and scalability, offer a robust and general-purpose foundation for our slicer. 

\subsubsection*{Challenge-2:Missing certain dependencies in \kn{Joern}'s CPG}
\label{sec:challenges:cpg}

While \kn{Joern} offers a solid foundation, its standard CPG representation presents practical limitations. Notably, it often lacks essential dependency relationships. For instance, crucial syntactic structural dependencies, such as links between `if' and `else' blocks or between parent and child code blocks, are sometimes absent. This deficiency can lead to incomplete code slices for certain warnings. This limitation, along with our proposed solution, is detailed in \autoref{sec:slicer}.



\subsubsection*{Challenge-3:Lack of efficient inter-file dependency extractors}
\label{sec:challenges:farf}

While whole-program analysis could provide complete context, it's impractical for large projects due to high time costs. For example, \kn{Joern} takes nearly an hour to export CPG for the Juliet dataset via whole-program analysis. Moreover, as SAST tools may generate many warnings, repeatedly performing such costly analysis per warning makes CI/CD integration infeasible. 
Therefore, an efficient method for quickly identifying warning-relevant files is essential to constrain slicing to relevant code without analyzing the entire project.

There are several existing file/code dependency analysis approaches, but none of them suit our needs. For instance, \citet{savic2014language}'s method targets Java class-level bytecode, thus lacking C/C++ support and file-level granularity. Similarly, \citet{molina2021automatic}'s approach, specific to VR applications, is not generalizable to diverse project types. 
~\citet{cpp_dependencies}'s \kn{cpp-dependencies} analyzes C++ include dependencies but cannot identify symbols declared via \codet{extern} or symbols whose definitions and implementations are separated.



\begin{figure*}[!t]
    \centering
    \includegraphics[width=0.95\linewidth]{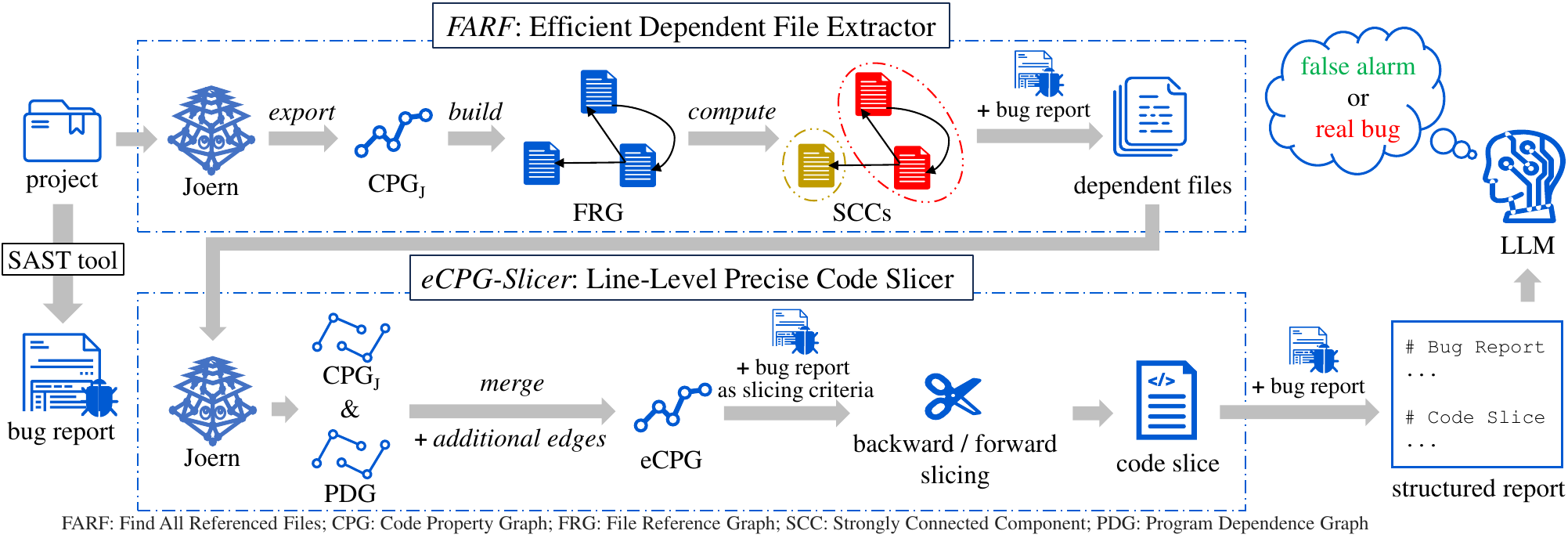}
    \vspace{-0.1cm}
    \caption{Overview of \mysys{}}
    \label{fig:mysys}
    \vspace{-0.1cm}
\end{figure*}

\section{Design}
\label{sec:design}


This section first introduces the design goal and overview of \mysys{} (\autoref{sec:design:overview}), then describes its key elements: the \slicer{} (\autoref{sec:slicer}), the \myalg{} algorithm (\autoref{sec:fdg}), and prompt engineering (\autoref{sec:llm_prompt}).


\subsection{Design goals and overview of \mysys{}}
\label{sec:design:overview}
\mysys{} aims to equip LLMs with precise and complete code context to enable accurate, reliable, and efficient FPM. To achieve this and address the challenges in \autoref{sec:bg}, we define the following design goals: 

\ding{172} \textbf{Precise Code Context.} \mysys{} must extract precise warning-relevant context from given source files while omitting extraneous details that could mislead the LLM. To address \hyperref[sec:challenges:slicer]{Challenge-1}, we implement a slicer based on \kn{Joern}. To tackle \hyperref[sec:challenges:cpg]{Challenge-2}, we enhance \kn{Joern}'s CPG by incorporating essential dependencies missing in original CPG. 

\ding{173} \textbf{Complete Code Context.} \mysys{} should account for key information often missing from bug reports, such as references to global variables or invoked functions located in other files. This goal involves providing the slicer with complete, warning-related input files. 

\ding{174} \textbf{High Efficiency.} As highlighted in \hyperref[sec:challenges:farf]{Challenge-3}, the efficient extraction of warning-related files is crucial for avoiding costly whole-program analysis. This goal, therefore, focuses on implementing an efficient algorithm to identify warning-related files. Such an approach will support the second goal and enhance the overall efficiency.


To achieve the above goals, we outline the entire process of \mysys{} in \autoref{fig:mysys}. It integrates our \slicer{} and \myalg{} algorithm. For each warning, the dependent file extractor applies \myalg{} to efficiently identify warning-relevant files. These source files, along with the corresponding bug report, are then passed to \slicer{} to generate precise line-level code context and form a structured report. This structured report is subsequently fed into an LLM using a carefully designed prompt, allowing it to analyze the report as a professional developer would and reach a final determination.

\slicer{} first merges \kn{Joern}'s CPG (hereafter $\mathrm{CPG_J}$) with PDG  before adding missing dependence edges to build an eCPG. This merging is necessary because, although the original CPG definition~\cite{yamaguchi2014modeling}---which we consolidate as \autoref{def:cpg}---includes the PDG, the $\mathrm{CPG_J}$ exported by \kn{Joern} lacks data dependency edges from PDG. For clarity, we refer to the CPG defined in \autoref{def:cpg} simply as ``CPG'' throughout the rest of the paper.

\begin{definition}[CPG]
\label{def:cpg}
    A CPG is defined as a graph \( G = (V, E, \lambda, \mu) \). \( V \) denotes a set of nodes given by AST nodes. \( E \subseteq (V \times V) \) represents a set of directed edges in AST, CFG, and PDG. \( \Sigma = \Sigma_{AST} \cup \Sigma_{CFG} \cup \Sigma_{PDG}\) is the set of edge labels involving labels from AST, CFG and PDG. \( \lambda: E \rightarrow \Sigma \) is an edge labeling function that assigns labels from \( \Sigma \) to edges. \( \mu: (V \cup E) \times K \rightarrow S \) is a property mapping, with \( K \) being a set of property keys (e.g., \codet{LINE\_NUMBER}) and \( S \) being a set of property values.
\end{definition}


\subsection{\slicer{} - a precise code context slicer based on eCPG}
\label{sec:slicer}

\begin{table*}[!t]
  \centering\footnotesize
  \caption{Construction rules for eCPG additional edges. $n_1 \xrightarrow{l} n_2$ denotes a directed edge from $n_1$ to $n_2$ with label $l$}
  \vspace{-0.1cm}
  \label{tab:edgeruls}
  \begin{tabular}{p{11cm}l} 
    \hline
    \bf Condition / Scenario & \bf eCPG Edge Construction Rule \\
    \hline 
    \multicolumn{2}{l}{\textbf{F-Edges (Call Dependence Edges, $E_F$)}} \\[1pt] \hline
    A call site node $c$ invokes function $f$ (with entry node $f_{\text{entry}}$). & 
    $E_F = E_F \cup \{ c \xrightarrow{F} f_{\text{entry}} \}$ \\ 
    $arg_i$ is the $i$-th actual argument node at call site $c$, and $param_i$ is the corresponding $i$-th formal parameter node of the callee function $f$. & 
    For each $i$: $E_F = E_F \cup \{ arg_i \xrightarrow{F} param_i \}$ \\ \hline

    \multicolumn{2}{l}{\textbf{S-Edges (Structural Dependence Edges, $E_S$)}} \\[1pt] \hline
    Node $n_1$ is an \texttt{if} statement and node $n_2$ is its true (\texttt{then}) block or corresponding \texttt{else} block. & 
    $E_S = E_S \cup \{ n_1 \xrightarrow{S} n_2 \}$ \\ 
    Node $n_1$ is a \texttt{switch} statement and node $n_2$ is one of its \texttt{case} or \texttt{default} blocks. & 
    $E_S = E_S \cup \{ n_1 \xrightarrow{S} n_2 \}$ \\ 
    Node $n_1$ is an outer code block and node $n_2$ is a nested sub-block for which $n_1$ defines the immediate enclosing scope. &
    $E_S = E_S \cup \{ n_1 \xrightarrow{S} n_2 \}$ \\ \hline

    \multicolumn{2}{l}{\textbf{V-Edges (Variable Dependence Edges, $E_V$)}} \\[1pt] \hline
    Node $decl$ is the declaration of a variable $v$, and node $use$ represents a usage of variable $v$. & 
    $E_V = E_V \cup \{ decl \xrightarrow{V} use \}$ \\
    \hline
  \end{tabular}
  \vspace{-0.1cm}
\end{table*}

\subsubsection{Limitation of CPG}
\label{sec:design:eCPG}



As noted in \hyperref[sec:challenges:cpg]{Challenge-2}, the PDG within CPG omits certain key dependence edges, limiting the slicer's ability to capture code context. 

\begin{figure}[htbp]
    \centering
    \includegraphics[width=0.95\linewidth]{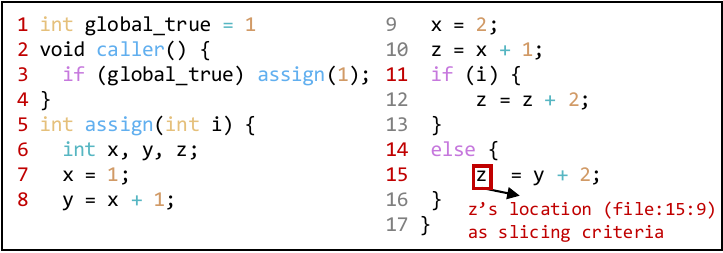}
    \vspace{-0.1cm}
    \caption{Example code. Given \codet{z} at line 15 as slicing criterion, red line numbers mark lines expected in the desired slice.}
  \label{fig:eg-slicer}
  \vspace{-0.1cm}
\end{figure}

Taking \autoref{fig:eg-slicer} as an example, if a warning involves variable \codet{z} and the slicing criterion is the triple (line 15, column 9, filename), the expected slice should include the code lines marked with red line numbers. These lines capture the data and control dependencies relevant to \codet{z}, such as how \codet{y} (line 8) depends on the value of \codet{x} (line 7), the condition determining which value is assigned to \codet{z} (line 11,14), and the calling context (line 3). Irrelevant code like \codet{z = x + 1;} (line 10) is removed. 
The desired slice precisely illustrates how \codet{z} (line 15) is derived, offering complete context related to its value. With this information, it becomes straightforward to trace \codet{z}'s data flow, allowing for an accurate determination of whether the warning is valid. 

However, relying solely on CPG's PDG results in an incomplete slice that includes only lines 15, 11, and 8, missing critical context such as the calling context of \codet{i} (line 3), the data flow into \codet{y} (line 7), and even the local declaration (line 6) within \kn{assign}.

\begin{figure}[htbp]
    \centering
    \includegraphics[width=0.95\linewidth]{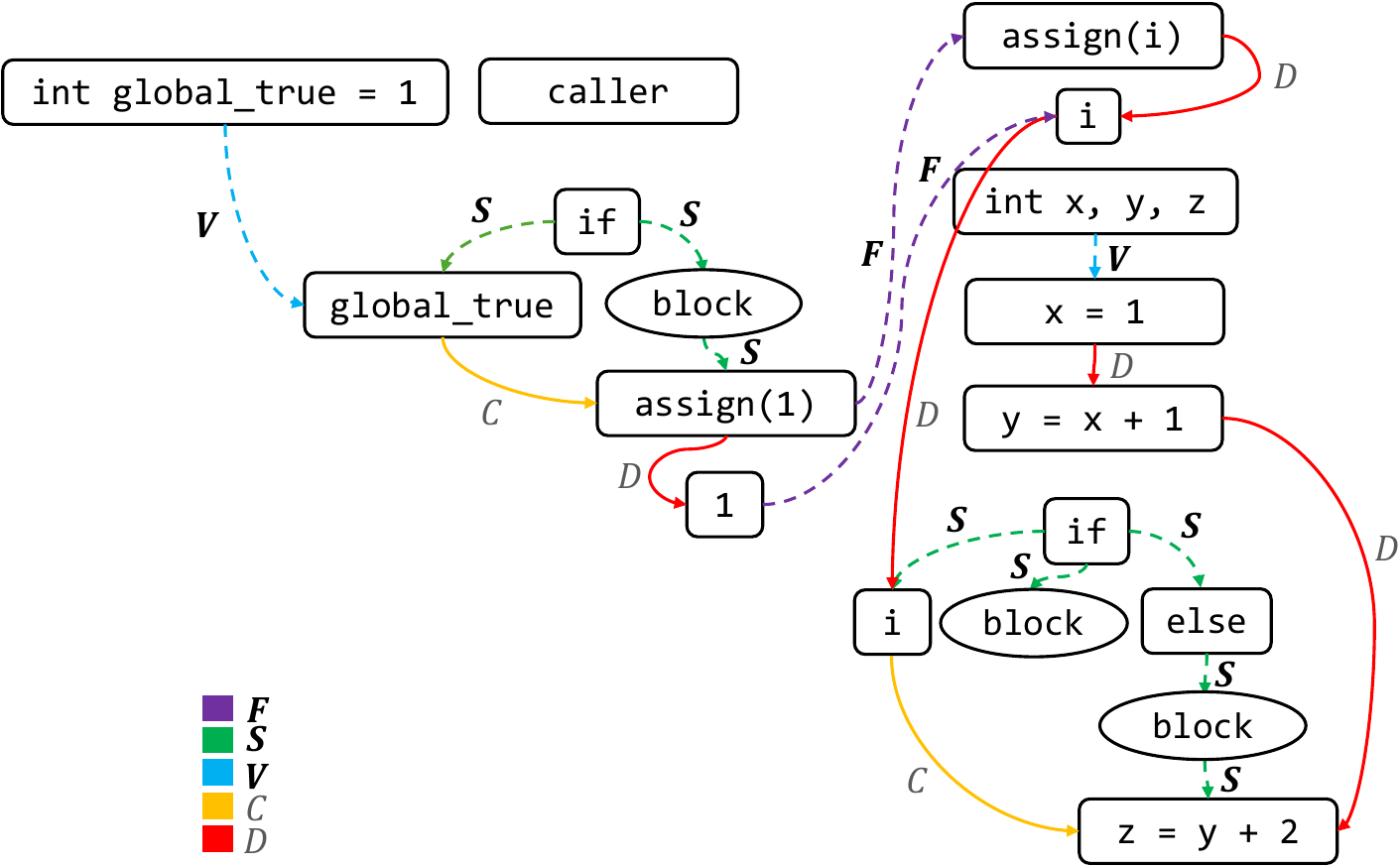}
    \caption{Simplified CPG and eCPG of \autoref{fig:eg-slicer}. $C$ and $D$ denote control and data dependence edges, respectively in CPG's PDG, indicated by solid lines. $F$, $S$ and $V$ denote call dependence, structural dependence and variable dependence edges respectively in our eCPG, indicated by dashed lines.}
    \label{fig:eCPG-example}
    \vspace{-0.1cm}
\end{figure}

\autoref{fig:eCPG-example} use solid lines to depict a simplified CPG derived from \autoref{fig:eg-slicer}. The imprecise slicing arises from the absence of three key relationship types in CPG’s PDG:
\\\ding{172} the \textbf{call dependence} between \codet{caller} and \codet{assign}, 
\\\ding{173} the \textbf{structural dependence} between \codet{if}, \codet{else}, and \codet{z=y+2}, 
\\\ding{174} the \textbf{variable dependence} between \codet{int x} and use of \codet{x}. 

\subsubsection{eCPG and its construction}

\autoref{def:pdg} extends the edge labeling set $\Sigma$ in CPG by introducing three new edge labels---$F$ (call dependence), $S$ (structural dependence), $V$ (variable dependence)---to capture the above dependencies.

\begin{definition}[extended dependence labels]
\label{def:pdg}
    \(\Sigma_{eCPG} = \{F, S, V, C, D\}\) where \(C\), \(D\) are the control dependence label and data dependence label from original $\Sigma_{PDG}$. \(F\), \(S\) and \(V\) are newly introduced, corresponding to the call, structural, and variable dependence, respectively.
\end{definition}

Let $E_{CPG}$ denote the edge set of CPG, and $E_{F}$, $E_{S}$ and $E_{V}$ denote the edge sets for new edge labels $F$, $S$ and $V$, 
then, our desired eCPG can be defined as \autoref{def:ecpg}.

\begin{definition}[eCPG]
\label{def:ecpg}
An eCPG is formally defined as a graph $G_{eCPG} = (V, E_{eCPG}, \lambda_{eCPG}, \mu)$. It builds upon the CPG defined in \autoref{def:cpg}, retaining the same set of nodes $V$ and the property mapping function $\mu$. Besides, $E_{eCPG} = E_{CPG} \cup E_{F} \cup E_{S} \cup E_{V}$. $\lambda_{eCPG}: E_{eCPG} \rightarrow \Sigma_{eCPG}$, assigns an appropriate label from $\Sigma_{eCPG}$ to each edge in $E_{eCPG}$.
\end{definition}

The key to constructing our eCPG is the creation of its $E_F$, $E_S$, and $E_V$ edges. In \autoref{tab:edgeruls}, we list the construction rules for these three newly introduced edge types. Specifically, Call Dependence Edges ($E_F$) link call sites to function entry points and map actual arguments to their corresponding formal parameters. Structural Dependence Edges ($E_S$) represent key syntactic relationships, including those linking if statements to else blocks, switch statements to their case or default blocks, and enclosing code blocks to their immediately nested sub-blocks. Finally, Variable Dependence Edges ($E_V$) connect variable declarations to their uses.

Applying these rules, we can construct an eCPG for the code example in \autoref{fig:eg-slicer}. A simplified depiction of this eCPG is presented in \autoref{fig:eCPG-example}, where dashed lines indicate the newly added edges. Specifically, the call dependence edges provide inter-procedural details by linking function calls (\codet{assign(1)} at line 3) to their definitions and mapping specific argument values (like \codet{i=1}), enabling LLMs to accurately trace execution paths and calling context. The dashed structural dependence edges make explicit the syntactic hierarchy, ensuring, for instance, that line 15 is correctly identified within its \codet{else} block, thus preventing LLM misinterpretation of code structure. Finally, the blue dashed variable dependence edges connect variable declarations (e.g., \codet{global\_true} on line 1 and local variables on line 6) to their usages, equipping LLMs with essential type and initialization information for more precise warning analysis.

\subsubsection{Slicing on eCPG}
\label{subsubsec:slicing}


Ottenstein and Ferrante et al.~\cite{10.1145/800020.808263, 10.1145/24039.24041} first propose that program slicing can be performed in linear time on graph representations of programs. 
We follow this idea and take the slicing procedure as resolving the reachability of a graph given certain vertices.


{
\begin{algorithm}[b]
    \caption{Generate source-level slice based on eCPG}
    \label{slicing-algoritm}
    \small
    \begin{algorithmic}[1]
        \REQUIRE program $P$ and its $eCPG$, a set of slicing criteria $C = \{(line, column, file name)\}$ 
        \ENSURE Slice for $P$ in terms of slicing criteria $C$    
        \STATE $N_{C} \gets$ nodes in $eCPG$ whose location in $C$
        \STATE Initialize worklist $W \gets N_{C}$, result set $R \gets \emptyset$
        \WHILE {$W \neq \emptyset$}
            \STATE $n \gets$ pop a front element from $W$
            \IF{$n$ is not visited}
            \STATE $R \gets R \cup \{n\}$
            \FOR{each incoming edge $e$ to $n$ with labelof($e$) $\in \Sigma_{eCPG}$}
                \STATE $W \gets W \cup $ \{source node of $e$\}
            \ENDFOR   
            \ENDIF
        \ENDWHILE
        \STATE Slice $S \gets$ ReconstructSourceSlice($R$, $\mu$) 
        \RETURN $S$
    \end{algorithmic}
    \vspace{-0.1cm}
\end{algorithm}
}

\autoref{slicing-algoritm} represents the complete slicing algorithm. Once eCPG has been constructed, \slicer{} finds corresponding nodes in terms of input slicing criteria. These criteria are triples of the form (line, column, filename) and extracted from bug reports. They specify distinct locations in the source code. Starting from these initial nodes, \slicer{} traverses the eCPG backwardly or forwardly to collect relevant nodes along dependence edges. The traversal process is implemented by a work-list algorithm. Finally, all collected nodes are mapped to source code via eCPG's property mapping function $\mu$, utilizing node properties(such as line numbers and filenames). This enables \slicer{} to output line-level slices while still maintaining the source-level information. 
Through the slicing algorithm, as shown in \autoref{fig:eCPG-example}, the nodes relavent to \codet{z} at line 15 in \autoref{fig:eg-slicer} are collected along dependence edges, which are mapped to source code in \autoref{fig:eg-slicer} at last.

\subsection{Dependent file extractor - efficient dependent file extraction}
\label{sec:fdg}
\paragraph*{Overview}

To efficiently identify all files relevant to each inspected warning, as necessitated by \hyperref[sec:challenges:farf]{Challenge-3}, our proposed two-step algorithm first performs a one-time, per-project analysis: creating a file reference graph (FRG) and computing its strongly connected components (SCCs). Subsequently, for each warning, these pre-calculated SCCs enable the rapid identification of all related files. These files are then provided to \slicer{}, thereby avoiding a costly whole-program analysis for \slicer{}.


\paragraph*{Step1: A one-time analysis to obtain FRG and SCCs} 


As shown in \autoref{fig:mysys}, first, we use Joern to perform a whole-program analysis on the project and export the project's $\mathrm{CPG_J}$. We then traverse the $\mathrm{CPG_J}$ to construct a directed File Reference Graph (FRG), where an edge from file A to file B signifies that A calls a function or references a variable defined in B. Subsequently, Tarjan's algorithm~\cite{tarjan1972depth} is applied to the FRG to compute its strongly connected components (SCCs). A condensation graph is then derived by contracting each SCC into a single vertex. Both the SCCs and the condensation graph are stored for repeated utilization in Step 2.

{
\begin{algorithm}[b]
    \caption{Find All Referenced Files(FARF)}
    \label{alg:find_referenced_files}
    \small
    \begin{algorithmic}[1]
        \REQUIRE FRG \( G_f \), SCCs \( S \), Condensation Graph \( G_{scc} \), File Set \( F \)
        \ENSURE Set of files \( F_{dep} \) that \( F \) depends on

        \STATE Initialize $F_{dep} \gets \emptyset$, $Visited \gets \emptyset$, $Queue \gets \emptyset$

        \FOR{each file $f$ in \( F \)}
            \STATE \( c \gets \text{the strongly connected component for } f \text{ in } S \)
            \IF{c not in visited}
                \STATE $Queue \gets Queue \ + $ $c$
                \STATE $Visited \gets Visited \ \cup $ \{$c$\}
            \ENDIF
        \ENDFOR

        \WHILE{$Queue \neq \emptyset$}
            \STATE $c_{cur} \gets$ pop a front element from $Queue$
            \STATE \( F_{dep} \gets F_{dep} \cup \{ \text{the files in } c_{cur} \text{ in } G_f \) \}
            \FOR{each neighbor $n$ in \( G_{scc} \text{ that } c_{cur} \text{ connects to } \)}
                \IF{neighbor  $n$ not in visited}
                    \STATE $Queue \gets Queue \ + $ $n$
                    \STATE $Visited \gets Visited \ \cup $ \{$n$\}
                \ENDIF
            \ENDFOR
        \ENDWHILE

        \RETURN \( F_{dep} \)
    \end{algorithmic}
\end{algorithm}
}

\paragraph*{Step 2: Use SCCs to efficiently obtain the files a warning depends on.} 

After obtaining the FRG, SCCs, and the condensation graph, \autoref{alg:find_referenced_files} identifies the complete set of files dependent on a given input set by traversing the condensation graph. Starting from the SCCs of the input files, it explores neighboring nodes to ensure all dependencies are comprehensively captured. The worst-case time complexity of \autoref{alg:find_referenced_files} is \(O(|V_{G_{f}}|)\), meeting our efficiency requirement for rapidly identifying warning-related files. Supplying these files to \slicer{} ensures complete context extraction without costly whole-program analysis.

\subsection{Guide the LLM}
\label{sec:llm_prompt}

After obtaining the precise and complete code context, we compile it together with the bug report generated by the SAST tool into a comprehensive, structured report that is then fed to the LLM. To effectively guide the LLM in providing accurate judgments, we adopt the following steps:

\textbf{1) Expert Prompting}. In the initial phase, we employ expert prompting~\cite{xu2023expertprompting} to guide the LLM to act as a proficient C/C++ warning reviewer. By emulating an expert's analytical process, the LLM can more effectively scrutinize the outputs of the SAST tool critically, thereby enhancing the accuracy of its assessments. 

\begin{figure}[htb]
    \vspace{-0.1cm}
    \centering
    \includegraphics[width=0.45\textwidth]{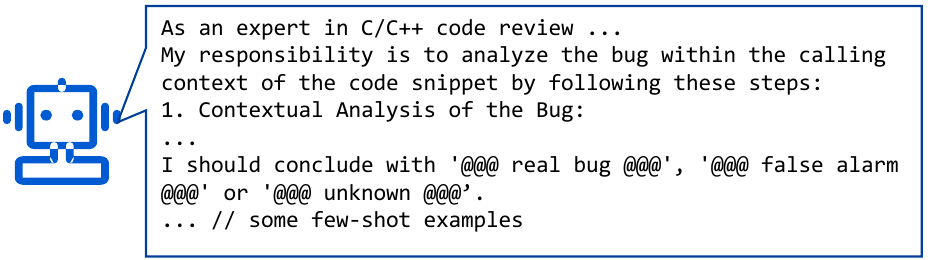}
    \vspace{-0.1cm}
    \caption{CWE121 buffer overflow prompt overview}
    \label{fig:prompt}
    \vspace{-0.1cm}
\end{figure}

\textbf{2) Chain-of-Thoughts Prompting}. 
We apply chain-of-thought prompting~\cite{wei2022chain} to guide the LLM in analyzing bug reports and corresponding code slices, requiring it to articulate its reasoning step by step for clarity and transparency. We also tailor prompting strategies to specific CWE types. For instance, as shown in \autoref{fig:prompt}, for buffer overflows, the LLM's output must include: \ding{172} Contextual Analysis of the Bug. \ding{173} Function Call Analysis and Parameter Correspondence. \ding{174} Conditional Judgment and Feasibility Analysis. \ding{175} Array Size and Index Analysis. \ding{176} Bug Verification.

\textbf{3) Few-Shot Prompting}. In addition to the chain-of-thought approach, we incorporate few-shot prompting~\cite{brown2020language} by providing a small set of illustrative question-answer examples. This technique allows the LLM to draw upon previous knowledge, enabling it to effectively address new scenarios. 



\textbf{4) Self-Consistency}. Finally, we integrate a self-consistency mechanism. Specifically, the LLM is configured to generate five responses, and a final conclusion is determined through a majority voting process. For each warning, the outcome(``false alarm'', ``real bug'' or ``unknown'') that appears most frequently is selected as the final decision. 



As an example, \autoref{fig:prompt} illustrates the prompt we prepared for CWE121. It highlights the key structures in the prompt, with some details omitted using ``...'' due to space constraints.



\section{Evaluation}
\label{sec:eval}
This section evaluates \mysys{} to demonstrate how  precise and complete code context positively influences LLM judgment. We will address the following research questions:

\noindent\textbullet\ \textbf{RQ1}: How does \mysys{} perform on existing datasets?\newline
\noindent\textbullet\ \textbf{RQ2}: How does \mysys{} perform on real-world projects?\newline
\noindent\textbullet\ \textbf{RQ3}: How does precise and complete code context affect LLM judgment?\newline
\noindent\textbullet\ \textbf{RQ4}: What is the cost of \mysys{}?

\subsection{Setup}
\label{sec:eval:conf}
We chose~\citet{wen2024automatically}'s LLM4SA discussed in \autoref{sec:bg}, which represents the SOTA as our baseline. To ensure consistency with the baseline, we selected the same three SOTA SAST tools for C/C++ projects: Cppcheck~\cite{cppcheck}, Infer~\cite{infer}, and CSA~\cite{csa}, and the same 7 representative categories of CWEs, including: \ding{172} CWE121 Stack-Based Buffer Overflow, \ding{173} CWE122 Heap-Based Buffer Overflow, \ding{174} CWE369 Divide by Zero, \ding{175} CWE401 Memory Leak, \ding{176} CWE416 Use After Free, \ding{177} CWE457 Use of Uninitialized Variable and \ding{178} CWE476 NULL Pointer Dereference.


To illustrate the cost-effectiveness of \mysys{} and to demonstrate that it does not rely on the model's inherent capabilities but instead uses precise and complete code context to guide the model toward correct conclusions, we utilize the open-source and free model \kw{qwen2.5-32b}~\cite{qwen2.5} instead of costly, large-scale proprietary models.
The model is deployed on a local server 
utilizing 4 RTX 3090 GPUs.

\subsection{RQ1 - How does \mysys{} perform on existing datasets?}
\label{sec:eval:rq1}

\begin{figure*}[htbp]
    \centering
    \includegraphics[width=0.85\linewidth]{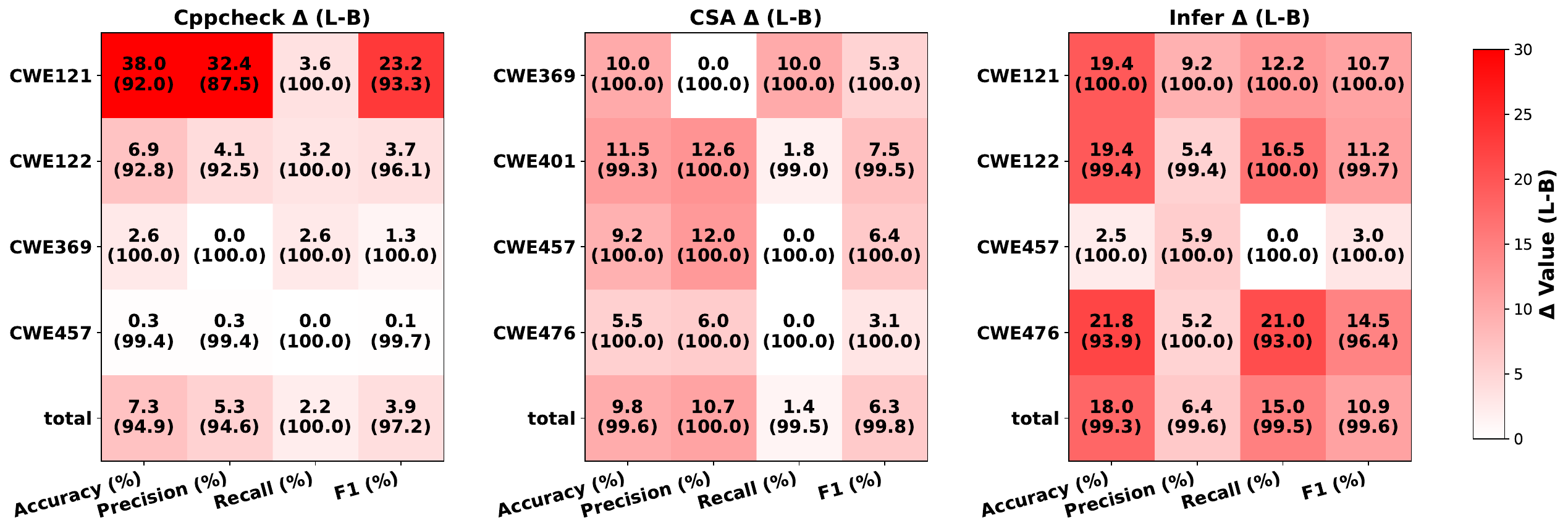}
    \vspace{-0.2cm}
    \caption{Delta heatmap of inspection performance metric differences between $L$ and $B$ for warnings generated by Cppcheck, CSA and Infer on the Juliet dataset. Each cell's first line displays the value of $L - B$ and second line (in parentheses) displays the value of $L$ for a given metric and CWE, with warmer colors indicating larger improvements of $L$ over $B$.}
    \label{fig:mysys-heatdelta}
    \vspace{-0.5cm}
\end{figure*}

For the datasets used in this RQ, ~\citet{croft2023data} comprehensively investigated 4 representative software vulnerability datasets. From these, we selected Juliet~\cite{juliet} and D2A~\cite{zheng2021d2a} because these two datasets are specifically designed for testing SAST tools. In contrast, the other two datasets are not intended for this purpose and, furthermore, their formats do not meet our requirements. The Juliet dataset, being synthetically generated and entirely accurate, is optimally suited for evaluation purposes; consequently, in this RQ, we employed Juliet to compare \mysys{} against the baseline. As for the D2A dataset, the labels are inaccurate, with a accuracy of approximately 53\%~\cite{zheng2021d2a}. Consequently, we chose not to use these original labels and tried to relabel the dataset with \mysys{} to enhance label accuracy in this RQ.


The Juliet dataset comprises a large number of C and C++ programs, including both benign cases and those exhibiting specific CWEs. SAST tools generate a large number of warnings in this labeled dataset, including false positives and true positives. For the 7 CWEs we consider, we obtain 7194 valid warnings in the Juliet dataset. 
For each warning in the Juliet dataset, we assess the LLM's performance. If code is genuinely buggy (a true SAST warning), an LLM confirmation is a TP; a dismissal or `unknown' response is an FN. If the SAST warning is a false positive (code not buggy), an LLM dismissal was a TN; an LLM confirmation of a bug or an `unknown' response is an FP. 
In the figures and tables below, $B$ represents the baseline, and $L$ represents \mysys{}.

\autoref{fig:mysys-heatdelta} presents a delta heatmap of the results, where each cell indicates $L$'s improvement over $B$ for a specific metric; warmer colors indicate greater improvements. For Cppcheck's CWE401, CWE416, and CWE476, both the baseline and \mysys{} achieve accuracy, precision, and recall of 100\%, so these results are excluded. It can be seen that for the warnings generated by Cppcheck, \mysys{} outperforms the baseline across all metrics, especially in CWE121. For CWE369 and CWE457, Cppcheck generates a limited number of warnings, and the code is simple, both baseline and \mysys{} perform well. 
For warnings generated by CSA and Infer, \mysys{} shows significant improvements over the baseline across each CWE. For the warnings generated by CSA, \mysys{} increases the F1 score from 93.48\% to 99.75\%. For the warnings from Infer, we improve the F1 score from 88.65\% to 99.57\%. These results demonstrate the effectiveness of \mysys{}. A key limitation of the baseline approach, explaining its underperformance against \mysys{} (especially for CWE121), is the lack of essential context like call chains and external variables needed to interpret warnings. This will be exemplified in RQ3.


The D2A dataset comprises warnings detected by Infer in real-world projects, annotated with binary labels indicating whether each warning is true. These labels are generated by an automated procedure and exhibit an overall accuracy of approximately 53\%~\cite{zheng2021d2a}. To improve the label accuracy, we used \mysys{} to relabel the data. Focusing on the libtiff subset, our approach successfully processed 881 instances. To validate the relabeling results, we randomly selected 50 cases for manual verification, conducted by two authors. Of these, we found that 43 cases matched the manual verification results, indicating that \mysys{} achieved an accuracy of 86\% on this sample. Upon analyzing the failed cases, we found three were due to the current inability of \kn{Joern} to expand code macros. This limitation prevents LLMs from observing variable assignments and references within the macro-expanded code, ultimately leading to incorrect results. The remaining four failures stemmed from the LLMs misinterpreting code semantics, even when provided with relevant context, potentially because such context was extensive. These findings indicate that ensuring correct semantic interpretation by LLMs remains a challenge in complex scenarios, even with precise and complete contextual information.



\begin{tcolorbox}[colframe=black, colback=gray!20, coltitle=gray!20, fonttitle=\bfseries, title=Answer to RQ1]
\vspace{-0.2cm}
Evaluation results on the Juliet dataset show that \mysys{} surpasses the existing baseline in all aspects. Furthermore, it successfully raises the D2A dataset’s accuracy to 82\%. The results across both datasets underscore the effectiveness of \mysys{}. Key factors constraining further performance enhancements include \kn{Joern}'s parsing limitations and LLM difficulties with complex context comprehension.
\vspace{-0.2cm}
\end{tcolorbox}

\subsection{RQ2: How does \mysys{} perform on real projects?}
\label{sec:eval:rq2}

In this RQ,  we evaluated the warning inspection capability of \mysys{} in real-world projects.


We used the GitHub API to find the 30 most-starred C/C++ repositories on GitHub. 
After cloning these repositories, we manually filtered out those with complex dependencies or compilation issues, ultimately selecting 8 projects with over 20k stars each. These projects vary in size from 30k to 590k lines of code, with an aggregate of 2.56 million lines. This constitutes a sufficiently large and representative sample. Then we analyzed these projects using SAST tools, including Cppcheck, Infer, and CSA. As shown in \autoref{tab:app-warnings}, the LOC denotes the lines of code, the Cppcheck, Infer, and CSA columns indicate the number of warnings detected by each tool; a ``-'' signifies that the tool failed to produce a result.

\vspace{-0.1cm}
\begin{table}[htbp]
\centering
\caption{Warnings detected on real-world projects}
\vspace{-0.1cm}
\label{tab:app-warnings}
\begin{tabular}{lrrrrr}
    \toprule
    \textbf{Projects} & \textbf{Stars} & \textbf{LOC}   & \textbf{Cppcheck} & \textbf{Infer} & \textbf{CSA} \\
    \midrule
    scrcpy     & 111k  &  29,898 &    0 &   – &   – \\
    netdata    & 71.7k & 591,633 &  126 &   0 &   0 \\
    redis      & 66.8k & 299,715 &    9 & 425 &  52 \\
    git        & 52.3k & 411,022 &    7 & 1130 &  60 \\
    curl       & 35.8k & 268,853 &   53 &  57 &   2 \\
    tmux       & 35.2k &  87,778 &    2 & 285 &  25 \\
    openssl    & 25.8k & 808,596 &   20 & 2105 & 122 \\
    masccan    & 23.5k &  60,733 &    5 & 180 &   4 \\
    \midrule
    \multicolumn{2}{l}{\textbf{total}} & 2,558,228 & 176 & 5255 & 323 \\
    \bottomrule
  \end{tabular}
\vspace{-0.2cm}
\end{table}

We ran \mysys{} on these warnings from 8 chosen projects. A random sample of 50 results was then manually inspected by 2 authors to assess its accuracy and overall effectiveness. The samples were selected multiple times to ensure coverage. 
Of the 50 samples we analyzed, the authors confirmed 43 samples as definitely false positives generated by SAST tools. The remaining could not be definitively categorized 
due to our lack of expertise with these repositories. Among the 43 samples we confirmed as false positives, \mysys{} successfully identified 37 cases as indeed false alarms generated by the SAST tool, achieving a success rate of 86.05\%. For the other six instances within this set of 43, \mysys{} produced one `unknown' response and incorrectly identified five as real bugs. We subsequently analyzed the underlying reasons for these six failures.

Four failures were attributed to \kn{Joern}’s inability in parsing certain C language features, such as code with extensive \codet{\#IFDEF} directives. This often resulted in incomplete CPG for specific files. This parsing limitation, in turn, prevented our \slicer{} from identifying all warning-related nodes, causing it to default to outputting the entire function containing the warning. Consequently, the provided code context became imprecise, leading the LLM to inaccurate conclusions. Another failed case stemmed from limitations in how \kn{Joern} and our \slicer{} identify function pointers. This issue prevented the detection of a specific function called via a function pointer in an \kn{openssl} example. The resulting omission of a crucial context segment caused the LLM to provide an incorrect response. In the final failed instance, \kn{Joern} and \slicer{} did not correctly determine if a specific variable was modified during a function call, leading to missing data dependency edges. As a result, the generated code snippet lacked a key function, rendering the code context incomplete. This absence of critical information prevented the LLM from ascertaining whether a specific variable had been initialized, ultimately causing an incorrect conclusion.

Lastly, and notably, even after slicing, some code slices still span hundreds or even thousands of lines, indicating that real-world code complexity can still be high despite slicing.

\begin{tcolorbox}[colframe=black, colback=gray!20, coltitle=gray!20, fonttitle=\bfseries, title=Answer to RQ2]
\vspace{-0.2cm}
Evaluation in RQ2 shows that \mysys{} reduces false positives by over 85\% in real-world projects. While issues such as function pointer dependencies and Joern's limitations may restrict full context recovery, the result remains a significant step forward. The inherent complexity of real-world code underscores the importance of effective information extraction.
\vspace{-0.2cm}
\end{tcolorbox}

\subsection{RQ3: How does precise and complete code context affect LLM judgment?}
\label{sec:eval:rq3}

In this RQ, we demonstrate the efficacy of \slicer{} and \myalg{}, the two core components in \mysys{}, through ablation studies (as detailed in \autoref{sec:rq3:ablation}). RQ3 is designed to ensure that the observed effectiveness genuinely stems from the specific contributions of \slicer{} and \myalg{}, rather than the LLM's inherent capabilities, particularly considering the possibility that the LLM may have been pre-trained on public datasets such as Juliet. Furthermore, we present a case study using a practical example from the Juliet dataset to illustrate how the complete and precise code context extracted by \mysys{} contributes to its effectiveness.

\subsubsection{Ablation Study}
\label{sec:rq3:ablation}

To validate the distinct contributions of the precise slices from \slicer{} and the complete file input from \myalg{}, we designed the following ablation experiments:

\begin{enumerate}
    \item \textbf{\mysys{}-noslice}: This variant bypasses the \slicer{}. Instead of slicing the code, the complete content of the files identified by the \myalg{} algorithm is directly fed to the LLM.
    \item \textbf{\mysys{}-nofarf}: This variant omits the \myalg{} algorithm. Code slicing is performed directly on the file explicitly indicated by the bug location in the bug report, without leveraging \myalg{} for relevant file identification.
\end{enumerate}

The ablation studies were performed on the Juliet dataset, employing the same experimental configurations as those for \mysys{} on this dataset. The ablation study results are shown in \autoref{tab:eval:ablation}. The full \mysys{} system achieves the best results (98.17\% Accuracy, 98.92\% F1). Performance significantly decreased when removing \myalg{} (\mysys{}-nofarf: 90.16\% Accuracy, 94.39\% F1), highlighting its importance for identifying related code files beyond the initially reported file. Excluding the \slicer{} (\mysys{}-noslice) also reduces performance (95.15\% Accuracy, 97.19\% F1), highlighting the slicer's effectiveness in focusing the LLM on relevant code segments. Crucially, \mysys{}-noslice additionally fails on 40 test cases because the input files exceeded the LLM's context window limit, demonstrating the slicer's practical necessity for managing input size and ensuring the feasibility of the approach. These findings underscore that both components are essential for the high effectiveness and applicability of \mysys{}. Next, we present an example from Juliet, correctly identified as a false positive by \mysys{} but missed by the baseline, to demonstrate how the precise and complete code context extracted by \mysys{} enables the LLM to recognize such false positives.


\begin{table}[htb]
\caption{Ablation study results (Notes: 40 cases exceed LLM's context limit in \mysys{}-noslice)}
\vspace{-0.1cm}
\label{tab:eval:ablation}
\scalebox{0.95}{
\begin{tabular}{lllllll}
\hline
 & \textbf{TP} & \textbf{TN} & \textbf{FP} & \textbf{FN} & \textbf{Accuracy} & \textbf{F1} \\ \hline
\mysys{}-nofarf & 5955 & 531 & 569 & 139 & 90.16\% & 94.39\% \\ 
\mysys{}-noslice & 6041 & 804 & 296 & 53 & 95.15\% & 97.19\% \\ 
\mysys{} & \textbf{6072} & \textbf{990} & \textbf{110} & \textbf{22} & \textbf{98.17\%} & \textbf{98.92\%} \\ \hline
\end{tabular}
}
\vspace{-0.1cm}
\end{table}

\subsubsection{Case Study}

\begin{figure}[htbp]
    \vspace{-0.1cm}
    \centering
    \includegraphics[width=0.45\textwidth]{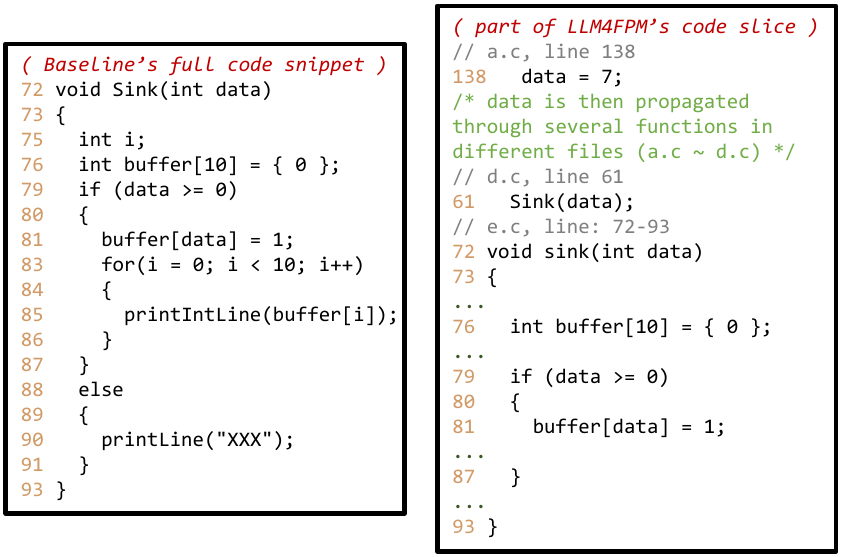}
    \caption{Simplified code snippet for the illustrative case. Green comments denote code segments omitted due to space constraints in \mysys{}'s code slice.}
    \label{fig:snippet}
\end{figure}
\begin{figure}[htbp]
    \centering
    \includegraphics[width=0.95\linewidth]{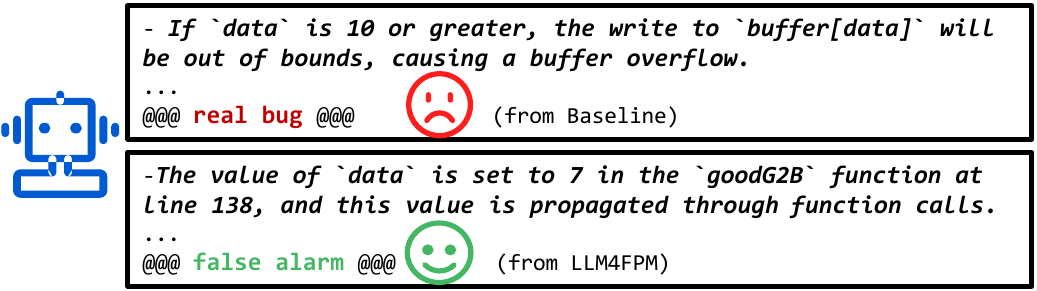}
    \caption{Output of Baseline and \mysys{}}
    \label{fig:output}
\end{figure}

Below, we use a false positive warning on Juliet as a typical case where \mysys{} successfully identified it as a false positive, while the baseline failed to do so. 
In this case, Cppcheck reports a false positive buffer overflow in line 81 of \autoref{fig:snippet}. The baseline method, by extracting only the function body where the warning occurs (i.e, the left part in \autoref{fig:output}), misled the LLMs into concluding that this is a real bug (as shown in the upper part of \autoref{fig:output}).
In contrast, our \mysys{} provided a precise and complete context (the right part in \autoref{fig:snippet}) by tracing the full multi-file call chain (from \codet{a.c} to \codet{e.c}). This reveals that the variable \codet{data} is assigned the value 7 in \codet{a.c} (line 138) and propagated to the warning site. Such comprehensive information ultimately guides the LLM to the correct conclusion that \codet{data} does not exceed 10, and consequently, that the warning is a false alarm (as depicted in the lower part of \autoref{fig:output}). 


This case serves as a concise illustration. Similar challenges in obtaining adequate context often arise from missing callee information or global variable definitions, details frequently omitted in SAST tool reports, or from overly verbose code that obscures essential control and data flows. In summary, this example underscores the positive impact of precise and complete code context on an LLM's analytical performance, which is also consistent with our ablation study results.

%
\begin{tcolorbox}[colframe=black, colback=gray!20, coltitle=gray!20, fonttitle=\bfseries, title=Answer to RQ3]
\vspace{-0.2cm}
Providing precise and complete code context guides LLMs to correct conclusions; conversely, flawed and incomplete code context can mislead them.
\vspace{-0.2cm}
\end{tcolorbox}

\subsection{RQ4: What is the cost of \mysys{}?}
\label{sec:eval:rq4}

In this RQ, we evaluated the costs of \mysys{}, including time cost and token usage. We estimated the reduced costs of \mysys{} compared to GPT-4.

Our time-cost testing, employing 12 concurrent requests with full GPU utilization on a local server, revealed an average of just 4.7 seconds per warning inspection. By comparison, manual inspection of a single bug can take several days~\cite{sawarkar2019predicting}. This low time overhead indicates integrating \mysys{} into CI/CD pipeline is feasible, allowing developers to focus on real bugs without being distracted by false positive warnings, contributing to improved software development quality.

For tokens, we counted both input tokens (estimated from the structured bug report provided to the LLM) and total tokens. Since our approach utilizes a locally deployed open-source model, it incurs no direct token consumption costs. Potential cost savings were estimated by calculating the hypothetical cost for these token counts using standard GPT-4 pricing: \$30.00 per million input tokens and \$60.00 per million output tokens~\cite{gptpricing}.

\begin{table}[htb]
\centering
\caption{Token usage and cost savings of \mysys{}}
\label{tab:eval:token}
\begin{tabular}{ccccccc}
\hline
\textbf{$\mathrm{T}_i$} & \textbf{$\mathrm{T}$} & \textbf{$\mathrm{C}$} & \textbf{$\mathrm{Avg_{T_i}}$} & \textbf{$\mathrm{Avg_{T}}$} & \textbf{$\mathrm{Avg_{C}}$} \\ \hline
6,561,004 & 49,261,720 & \$2,758.87 & 912.01 & 6,847.61 & \$0.384 \\ \hline
\end{tabular}
\end{table}

Results on Juliet (\autoref{tab:eval:token}) indicate that for the entire dataset, \mysys{} used a total of 6,561,004 input tokens ($\mathrm{T}_i$) and 49,261,720 total tokens ($\mathrm{T}_i$). On average, this translated to approximately 912 input tokens ($\mathrm{Avg_{T_i}}$) and 6847 total tokens ($\mathrm{Avg_{T}}$, for 5 responses) per warning. This efficiency translates to an average cost saving of \$0.384 per warning ($\mathrm{Avg_{C}}$) leading to a total estimated saving of around \$2758 ($\mathrm{C}$) for the entire dataset. Considering SAST tools often generate thousands of warnings in large real-world projects, these results underscore \mysys{}'s potential for substantial cost reductions in practical deployments.

\begin{tcolorbox}[colframe=black, colback=gray!20, coltitle=gray!20, fonttitle=\bfseries, title=Answer to RQ4]
\vspace{-0.2cm}
\mysys{}'s low time overhead suggests that it can be integrated into the development process. \mysys{} saves thousands of dollars each run by utilizing the open-source model, making it a economical choice.
\vspace{-0.2cm}
\end{tcolorbox}

\section{Limitations and Future Work}
\label{sec:limits}

While \mysys{} demonstrates notable capabilities, several challenges remain. For \slicer{}, \ding{172} it can be impacted by current imperfections in Joern's parsing (e.g., issues with function pointers, incorrect attributes, or macros), which can affect slice accuracy. Addressing such imperfections necessitates further advancements in underlying program analysis frameworks and techniques. \ding{173} The current forward/backward slicing method could be evolved into more adaptive, fine-grained strategies. \ding{174} The processing overhead for CPGs from very large projects, a factor mitigated by \myalg{}, also warrants ongoing optimization efforts. \ding{175} Despite slicing, real-world warning's code slices often remain complex, potentially hindering the LLM's semantic comprehension of the code. Investigating how to effectively extract effective information from such complex code requires further study.


Regarding \myalg{}, its current file-level dependency granularity offers an area for refinement; moving to function or variable-level dependencies could yield greater precision, presenting a trade-off with potential overhead to be explored.

Finally, for the \mysys{} framework itself, handling warnings that appear in multiple code contexts without specific disambiguation from SAST tool reports can sometimes limit the precision of the extracted context. Addressing these limitations will be the focus of future efforts to further enhance the robustness, precision, and broader applicability of \mysys{}.

\section{Related Work}
\label{sec:related}

During the past years, many FPM approaches have been proposed. Some studies employ program analysis techniques; for instance, \citet{giet2019towards, muske2019reducing, liang2012effective} utilized dependency flow analysis, while \citet{rival2005understanding, rival2005abstract} applied code slicing methods. Researchers have also explored AI-based solutions for FPM. Before the advent of LLMs, early efforts often involved traditional machine learning: \citet{koc2017learning}, for example, used such techniques on reduced code to learn a classifier for filtering false positives, and \citet{ranking2014finding} trained models to identify actionable alerts.
The advent of LLMs, with their powerful capabilities in natural language processing and code understanding, brings new potential for programming and code safety/security. For example, \citet{li2023poster} investigates how LLMs can assist static analysis by using carefully constructed questions to prune false positives. Other works highlight LLM proficiency in areas such as function-level fault localization~\cite{wu2023large}, precisely locating buggy code lines and vulnerabilities~\cite{yang2024large}, fixing synthetic bugs~\cite{pearce2023examining}, identifying complex use-before-initialization bugs~\cite{li2024enhancing}, and enhancing code review and understanding~\cite{yu2024fine}. 
\citet{wen2024automatically} were the first to propose a complete system, LLM4SA, to use LLMs to help determine whether a warning is a false alarm, which was the SOTA. In this paper, we identify its limitations and propose our \mysys{} framework, which surpasses LLM4SA in experiments.

 

\section{Conclusion}
\label{sec:conclusion}

This paper identifies a key gap in existing FPM approaches: the lack of precise and complete code context, which can mislead LLMs and result in incorrect judgments. By addressing this gap, our work reduces false positives produced by SAST tools. Specifically, we propose the \mysys{} framework, which extracts precise and complete code context to guide LLMs in mitigating false positives. \mysys{} leverages the \slicer{} and the \myalg{} algorithm to ensure high-quality context extraction. Experiments on the Juliet and D2A datasets, as well as real-world projects, show that \mysys{} is both effective and low-cost. These results underscore the value of precise code context for LLM performance and highlight the importance of extracting relevant information from complex real-world projects. Despite certain limitations such as the inadequacies of current program analysis tools and challenges in LLM comprehension of lengthy code semantics, our work highlights the promise of combining program analysis with LLMs to advance automated, efficient, and reliable software engineering.




\bibliographystyle{IEEEtranN}
\bibliography{sp}

\end{document}